\documentclass[aps,seceqnum,preprintnumbers,nofootinbib,superscriptaddress]{revtex4}
\usepackage[dvipdfmx]{graphicx}
\usepackage{amsmath, amssymb}
\usepackage{comment}
\usepackage{float}

\makeatletter
    
    \@addtoreset{equation}{section}
\makeatother

\begin{document}

\preprint{YITP-15-37, KUNS-2508}

\title{
Appearance of Boulware-Deser 
ghost in bigravity with doubly coupled matter
}

\author{Yasuho Yamashita}
\affiliation{Yukawa Institute for Theoretical Physics,
Kyoto University, 606-8502, Kyoto, Japan}

\author{Antonio De Felice}
\affiliation{Yukawa Institute for Theoretical Physics,
Kyoto University, 606-8502, Kyoto, Japan}

\author{Takahiro Tanaka}

\affiliation{Yukawa Institute for Theoretical Physics,
Kyoto University, 606-8502, Kyoto, Japan}
\affiliation{Department of Physics, Kyoto University, 606-8502, Kyoto, Japan}

\begin{abstract}
We discuss the ghost freeness in the case when we add matter coupled to two metrics 
to the ghost-free bigravity.
In this paper we show that the Boulware-Deser 
ghost generally revives 
in the presence of doubly coupled matter and that ghost freeness strongly restricts 
the model of kinetically doubly coupled matter.
This result may anticipate difficulties in the attempt to derive the ghost-free bigravity 
as a low-energy effective theory, starting with a model applicable at high energies.
\end{abstract}

\maketitle

\section{Introduction}

Bigravity, the theory that has two dynamical metrics interacting with
each other through non-derivative interaction terms, has been
investigated since the proposal in Ref.~\cite{Isham:1971gm}.  In the
presence of the mass of graviton which derives from the non-derivative
interaction terms, however, it is known that an additional degree of
freedom generally appears and that it has a wrong-sign kinetic term.
Such degree of freedom causes instability at the quantum level and is
called the Boulware-Deser (BD) ghost~\cite{BD}.  Recently, de Rham,
Gabadadze and Tolley discovered the unique ghost-free\footnote{
Here and in the
following ``ghost-free'' strictly means that the model does not have the BD ghost, without excluding the possible existence
of another type of ghost.}
 non-derivative interaction in
the case with one non-dynamical fiducial metric, which is called dRGT
massive gravity~\cite{dRGT1, dRGT2} (see also \cite{massiveG}). 
References~\cite{HR1, HR2, Hassan:2011ea} showed that dRGT massive gravity
is certainly ghost free and that the proof of ghost-freeness can be
extended to the case where both metrics are dynamical.  
  Reference~\cite{HR1} also argued that the case in which a matter field
  couples to both metrics is
  allowed, although such a matter field would in general
  violate the equivalence principle.

This finding makes it possible to study healthy bimetric theory
\cite{DeFelice:2013nba}. 
Many works about the ghost-free bigravity have been done, for example, 
in the cosmological context 
for the case where each matter couples only to either metric
~\cite{Comelli:2011zm, Comelli:2012db, Volkov:2012cf, Fasiello:2013woa,DeFelice:2014nja,Konnig:2014xva} 
(see also \cite{otherBG}) 
and for the case where matter couples to both metrics~\cite{Akrami:2013ffa, Akrami:2014lja,Tamanini:2013xia}.
In Refs.~\cite{Comelli:2012db, DeFelice:2014nja,Konnig:2014xva} 
they studied the nature of the cosmological solutions of the ghost-free bigravity 
and found that even in the healthy branch where Higuchi ghost is absent, 
a gradient instability appears in the high energy regime 
compared with the mass scale introduced in the Lagrangian 
in the matter-dominated or radiation-dominated era. 
Therefore, if the ghost-free bigravity can describe the real universe, 
it should exceed the range of validity in the high energy regime 
and it must be extended to some more fundamental theory there.
Such an extension of the ghost-free bigravity is attempted 
in Ref.~\cite{deRham:2013awa}, where the ghost free multigravity is related to 
higher dimensional general relativity with a discrete extra dimension, 
and in Ref.~\cite{Yamashita:2014cra}, 
where the ghost-free bigravity is embedded in the DGP 2-brane model.
In this context, we can naturally consider matter that couples to both metrics 
by introducing a five-dimensional matter field.

Despite the above discussion, 
in this paper we claim that the BD ghost generally appears 
when we add matter coupled to both metrics by hand to the ghost-free bigravity.
This is because, when the matter has two kinetic terms which couple to respective metrics,
the conjugate momentum of the matter field depends on both lapse functions and 
the kinetic term written by the conjugate momentum becomes nonlinear in two lapse functions.
Considering perturbations on Friedmann and anisotropic background solutions, 
we investigate in which case an extra degree of freedom is present 
and whether the extra degree of freedom is a ghost mode or not.
Consequently, when matter has two kinetic terms corresponding to two metrics, 
we find that an extra degree of freedom appears 
and that there is no mechanism to avoid the mode to become a ghost mode 
in the limit of Minkowski solution in the healthy branch. 

This paper is organized as follows.
In Sec.~2 we present a brief summary and a detailed follow-up calculation of the proof 
that has been given 
by Hassan and Rosen~\cite{HR1, HR2, Hassan:2011ea}. 
Here we consider only the gravitational sector and 
confirm that the ghost-free bigravity really has four Lagrange multipliers that remain unspecified 
by the consistency conditions of the time evolution of the constraints, 
corresponding to the gauge degrees of freedom, 
which was not explicitly shown in Refs.~\cite{HR1, Hassan:2011ea} 
by the calculation of the Poisson brackets 
among the Hamiltonian and momentum constraints. 
In Sec.~3, we consider the coupling to matter and investigate 
in which case the BD ghost appears in doubly coupled matter models. 
Section.~4 is devoted to the summary of this paper.

\section{gravitational sector}

In this section we give detailed calculations necessary in prooving that 
the Boulware-Deser ghost is really absent in the so-called ghost-free bigravity. 
First, we introduce the setup of the model following the notation 
in Refs.~\cite{HR1, HR2, Hassan:2011ea} 
and give a brief summary of the proof given in the above References.
The action that we consider is given by
\begin{align}\label{action}
 S= \int d^4x\sqrt{-{\rm det}\, g}\left[ \frac{M_g^2}{2}R^{(g)} 
 + 2m^2 M_{\rm{eff}}^2 
\sum_{n=0}^{4} c_n e_n \left(\sqrt{g^{-1}f} \right) \right] 
 + \frac{M_f^2}{2}\int d^4x\sqrt{-{\rm det}f} R^{(f)}\, ,
\end{align}
where $g_{\mu\nu}$ and $f_{\mu\nu}$ are, respectively, the
physical and the hidden metrics.
$M_g$ is the 4 dimensional Planck mass for $g_{\mu\nu}$, 
$M_f$ is that for $f_{\mu\nu}$, and $M_{\rm{eff}}$ is defined as
\begin{align}
   M_{\rm{eff}}^2 = \left( \frac{1}{M_g^2} + \frac{1}{M_f^2} \right)^{-1} \, .
\end{align}
We also introduce $m$ and $c_n$ (corresponding to $\beta_n$ 
in Ref.~\cite{HR1, HR2, Hassan:2011ea}) as model parameters. 
The interaction terms between $g_{\mu\nu}$ and $f_{\mu\nu}$ are given by
\begin{align}
   e_0 &= 1 \, , \ \ 
   e_1 = [Y] \, , \ \ 
   e_2 = [Y]^2 - [Y^2] \, , \ \ 
   e_3 = [Y]^3 - 3 [Y] [Y^2] + 2 [Y^3] \, , \nonumber \\
   e_4 &= [Y]^4 -6 [Y]^2 [Y^2] + 8 [Y] [Y^3] +3 [Y^2]^2 -6 [Y^4] \, ,
\end{align}
where we have introduced $Y^{\mu}_{\nu} = \sqrt{ g^{\mu\alpha} f_{\alpha\nu} } $ 
and $[Y^n] = \mathrm{Tr}(Y^n)$.

In this paper we count the number of degrees of freedom 
in the ghost-free bigravity defined by Eq.~\eqref{action} using Hamiltonian formulation.
To do Hamiltonian analysis, we decompose the metrics 
discriminating temporal and spatial components as
\begin{align}
   N^{-2} = -g^{00}\,,\ \ \ N_i = g_{0i}\,, \ \ \ \gamma_{ij} = g_{ij}\,, \nonumber \\
   L^{-2} = -f^{00}\,,\ \ \ L_i = f_{0i}\,, \ \ \ {}^3\!f_{ij} = f_{ij}\,.  
\end{align}
Using these variables, the Lagrangian of the ghost-free bigravity \eqref{action} becomes 
\begin{align}\label{Lag}
   \mathcal{L} = M_g^2 \left[ \pi^{ij} \partial_t \gamma_{ij} + NR^0_{(g)} + N_i R^i_{(g)} \right]
   + 2m^2 M_{\rm{eff}}^2 \sqrt{\rm{det}\gamma}\, N \sum_{n=0}^{4} c_n e_n \left(\sqrt{g^{-1}f} \right) \nonumber \\ 
   + M_f^2 \left[ p^{ij} \partial_t \,{}^3\! f_{ij} + LR^0_{(f)} + L_i R^i_{(f)} \right]\, ,
\end{align}
up to irrelevant total derivative terms, 
where $\pi^{ij}$ and $p^{ij}$ are the momenta canonically conjugate 
to $\gamma_{ij}$ and ${}^3f_{ij}$, respectively, and 
\begin{align}
   &R_0^{(g)} = \sqrt{\rm{det}\gamma}\, {}^3\! R_{(\gamma)} + \frac{1}{\sqrt{\rm{det}\gamma}} 
   \left( \frac{1}{2} \pi^i_i \pi^j_j - \pi^{ij} \pi_{ij} \right)\, , \\
   &R_i^{(g)} = 2 \sqrt{\rm{det}\gamma} \,\gamma_{ij} \nabla^{(g)}_k 
   \left( \frac{\pi^{jk}}{\sqrt{\rm{det}\gamma}} \right)\, . 
\end{align}
Here $\nabla^{(g)}_i$ and ${}^3\! R_{(\gamma)}$ are the covariant differenciation 
and the Ricci scalar defined by $\gamma_{ij}$, respectively.
$R_0^{(f)}$ and $R_i^{(f)}$ are defined in the same way as $R_0^{(g)}$ and $R_i^{(g)}$ 
using ${}^3\! f_{ij}$ and $p^{ij}$ instead of $\gamma_{ij}$ and $\pi^{ij}$.

Obviously this Lagrangian is nonlinear in terms of the lapse functions $N$ and $L$ 
and the shift vectors $N^i$ and $L^i$, and it is difficult to count 
the number of degrees of freedom using these variables.
In order to make the Lagrangian linear in $N$, $L$ and $L^i$, Hassan and
Rosen~\cite{HR1}
defined a new shift-like vector $n^i$ as
\begin{align}\label{n}
   N^i -L^i = \left( L \delta^i_j + N D^i_j \right) n^j\, ,
\end{align}
where $N^i = \gamma^{ij} N_j$, $L^i = {}^3\! f^{ij} L_j$ 
and the matrix $D$ is determined by the condition,
\begin{align}\label{D}
   \sqrt{x}D = \sqrt{\left( \gamma^{-1} - Dnn^T D^T \right) {}^3\! f}\, ,
\end{align}
where,
\begin{align}
   x := 1 - n^i \,{}^3\! f_{ij} n^j\, .
\end{align}
The condition that determines $D$ imposes an important property, 
\begin{align}
   f_{ik} D^k_j = f_{jk} D^k_i\, .
\end{align}
We often use this property in the following calculation without mentioning it.

Rewriting the shift vector $N^i$ using $n^i$, the Lagrangian becomes 
\begin{align}\label{Lagn}
   \mathcal{L} = M_g^2 \pi^{ij}\partial_t \gamma_{ij} + M_f^2 p^{ij} \partial_t \,{}^3\! f_{ij} 
   -\mathcal{H}_0 \, ,
\end{align}
where
\begin{align}
   \mathcal{H}_0 :=& - L^i \left( M_g^2 R_i^{(g)} + M_f^2 R_i^{(f)} \right) 
   - L \left( M_f^2 R^{0(f)} + M_g^2 n^i R_i^{(g)} 
   + 2m^2 M_{\rm{eff}}^2 \sqrt{{\rm det}\, \gamma}\, U \right)\, \nonumber \\
   &-N \left( M_g^2 R^{0(g)} + M_g^2 R_i^{(g)} D^i_j n^j
    +  2m^2 M_{\rm{eff}}^2 \sqrt{{\rm det}\, \gamma}\, V \right) \, .
\end{align}
$U$  and $V$ are defined as 
\begin{align}
   &U := c_1 \sqrt{x} + c_2 \left[ \sqrt{x}^2 D^i_i + n^i \, {}^3\!f_{ij} D^j_k n^k \right] \nonumber \\
   &\ \ \ \ \ + c_3 \left[ \sqrt{x} \left( D^l_l n^i \, {}^3\! f_{ij} D^j_k n^k 
   - D^i_l n^l \, {}^3\! f_{ij} D^j_k n^k \right) 
   + \frac{1}{2} \sqrt{x}^3 \left( D^i_i D^j_j - D^i_j D^j_i \right) \right] 
   + c_4 \frac{ \sqrt{ {\rm det}\, {}^3\! f} }{ \sqrt{ {\rm det}\, \gamma} }\, , \\
   &V := c_0 + c_1 \sqrt{x} D^i_i + \frac{1}{2} c_2 \sqrt{x}^2 
   \left[ D^i_i D^j_j - D^i_j D^j_i \right]+ \frac{1}{6} c_3 \sqrt{x}^3 
   \left[ D^i_i D^j_j D^k_k - 3D^i_i D^j_k D^k_j + 2 D^i_j D^j_k D^k_i \right]\, .
\end{align}
Variation of the Lagrangian~\eqref{Lagn} with respect to $n^k$ leads to the equations of motion, 
which determines $n^k$.
At this point the Lagrangian is already linear in $N$, $L$ and $L^i$.
Calculating the derivatives of $\mathcal{H}_0$ and $\mathcal{C}$ with respect to $n^k$, we obtain 
\begin{align}\label{CL/ni}
   &\frac{\partial \mathcal{H}_0}{\partial n^k} = L \mathcal{C}_k \, , \\
   \label{C/ni}
   &\frac{\partial \mathcal{C}}{\partial n^k} 
   = \mathcal{C}_i \frac{\partial \left( D^i_j n^j \right)}{ \partial n^k }\, ,
\end{align}
where 
\begin{align}\label{Ci}
   \mathcal{C}_i :=& R_i^{(g)} - 2\tilde{m}^2 \sqrt{ {\rm det}\, \gamma}\, \frac{n^l \,{}^3\! f_{lj}}{\sqrt{x}} \tilde{V}^j_i\, , \\
  \tilde{V}^j_i :=& \frac{1}{\sqrt{x}} \frac{\partial V}{\partial D^i_j} \nonumber \\
   =& c_1 \delta^j_i + c_2 \sqrt{x} \left[ \delta^j_i D^k_k - D^j_i \right] 
   + c_3 \sqrt{x}^2 \left[ \frac{1}{2} \left( D^m_m D^n_n - D^m_n D^n_m \right) \delta^j_i
   + D^j_k D^k_i - D^j_i D^l_l \right] \, ,
\end{align}
with $M_g^2 \tilde{m}^2 := M_{\rm eff}^2 m^2$
($\tilde{V}^j_i$ is defined differently in \cite{HR1}).
Here we used 
\begin{align}
\label{Vn}
   &\frac{\partial \left( \sqrt{\rm det\, \gamma} \,V \right)}{\partial n^k} 
   = -\frac{\sqrt{\rm det\, \gamma} }{\sqrt{x}} n^l \tilde{V}^m_l \,{}^3\! f_{mn} 
   \frac{\partial \left( D^n_j n^j \right)}{\partial n^k}\, ,  \\
   \label{Un}
   &\frac{\partial \left( \sqrt{\rm det\, \gamma} \,U \right)}{\partial n^k} 
   = - \frac{\sqrt{\rm det\, \gamma}}{\sqrt{x}} \tilde{V}^i_k \,{}^3\! f_{ij} n^j\, .
\end{align}
Therefore the equation of motion of $n^k$ becomes 
\begin{align}\label{EOMn}
   \frac{\partial \mathcal{L}}{\partial n^k} = \mathcal{C}_i 
   \left[ L \delta^i_k + N \frac{\partial \left( D^i_j n^j \right)}{ \partial n^k } \right]\, .
\end{align}
The matrix in the square brackets in Eq.~\eqref{EOMn}, which is the Jacobian of 
the transformation \eqref{n}, is invertible in general.
Therefore the equations of motion of $n^i$ become
\begin{align}\label{EOMni}
   \mathcal{C}_i = 0\, .
\end{align} 
These are independent of $N$, $L$ and $L^i$ (and $p^{ij}$), and hence we can solve 
these equations to describe $n^i$ in terms of $\gamma_{ij}$, $\pi^{ij}$ and ${}^3\!f_{ij}$.
Substituting $n^i \left( \gamma_{ij}, \pi^{ij}, {}^3\!f_{ij} \right)$ to the Lagrangian \eqref{Lagn},
we find that the Lagrangian is still linear in $N$, $L$, and $L^i$.
We obtain three primary constraints 
\begin{align}
    \label{C}
   &\mathcal{C} = M_g^2 R^{0(g)} + M_g^2 R_i^{(g)} D^i_j n^j
    +  2m^2 M_{\rm{eff}}^2 \sqrt{{\rm det}\, \gamma}\, V\, , \\
   \label{CL}
   &C^L = M_f^2 R_0^{(f)} + M_g^2 n^k R_k^{(g)} 
   + 2m^2 M_{\rm eff}^2 \sqrt{{\rm det}\, \gamma}\, U\, , \\
   \label{CLi}
   &C^L_i = M_g^2 R_i^{(g)} + M_f^2 R_i^{(f)}\, ,
\end{align} 
by taking the variations of this Lagrangian with respect to $N$, $L$, and $L^i$.
With Eqs~\eqref{EOMni}, Eqs.~\eqref{CL/ni} and \eqref{C/ni} imply 
that the constraints $\mathcal{C}$ and $C^L$ vanish 
under the differentiation with respect to $n^i$, and hence we can treat $n^i$ as if it were fixed 
when we calculate the variations of the primary constraints.

In Ref.~\cite{HR1}, it was claimed that in the case of dRGT massive gravity, 
where $f_{\mu\nu}$ is fixed and non-dynamical, 
the variation of the Lagrangian with respect to $N$ gives
a constraint $\mathcal{C} = 0$ and its consistency condition gives another constraint, 
whose consistency condition determines the Lagrange multiplier $N$. 
Hence, there are two constraints and Boulware-Deser ghost is absent. 
They also claimed that in the bigravity case this procedure can be simply extended and 
these two constraints, along with four constraints $C^L$ and $C^L_i$
produced by the variation with respect to $L$ and $L^i$, respectively,  
and four gauge fixing conditions, reduce 24 variables, 
$\gamma_{ij}$, $\pi^{ij}$, ${}^3\!f_{ij}$ and $p^{ij}$, to 14 degrees of freedom.
This counting is correct, but it is not confirmed 
whether all the Poisson brackets of the primary constraints really vanish 
except for the one between $\mathcal{C}$ and $C^L$. 
Although this might be almost obvious due to the existence of 
general coordinate invariance in the ghost-free bigravity, 
we explicitly calculate the Poisson brackets between the primary constraints 
to confirm that the proof of the absence of ghost in dRGT massive gravity 
can be extended to bigravity.

In order to calculate the Poisson brackets, we have to calculate 
the partial differentiations of $U$ and $V$ with respect to $\gamma_{ij}$ and ${}^3\!f_{ij}$ 
as well as $n^i$.
For this purpose, the following relations 
as given in Eqs.~\eqref{Vn} and \eqref{Un} for $n^i$ are useful:
\begin{align}
   \label{Vg}
   &\frac{\partial \left( \sqrt{\rm det\, \gamma} \,V \right)}{\partial \gamma_{ab}} 
   =  -\frac{\sqrt{\rm det\, \gamma} }{\sqrt{x}} \left[ n^l \tilde{V}^m_l \,{}^3\! f_{mn} 
   \frac{\partial \left( D^n_j n^j \right)}{\partial \gamma_{ab}} 
   -\frac{1}{2} \left( \sqrt{x} V \gamma^{ab} - \gamma^{al} (D^{-1})^m_l 
   \tilde{V}^n_m \,{}^3\! f_{nk} \gamma^{kb} \right)  \right]\, , \\
   \label{Vf}
   &\frac{\partial \left( \sqrt{\rm det\, \gamma} \,V \right)}{\partial \,{}^3\! f_{ab}} 
   = -\frac{\sqrt{\rm det\, \gamma} }{\sqrt{x}} 
   \left[ n^l \tilde{V}^m_l \,{}^3\! f_{mn} \frac{\partial \left( D^n_j n^j \right)}{\partial \,{}^3\! f_{ab}}
   -\frac{1}{2} x \tilde{V}^a_l \,{}^3\! f^{lm} D^b_m \right]\, , \\
   \label{gnU}
   &n^k \gamma_{ik} \frac{\partial \left( \sqrt{\rm det\, \gamma} \,U \right)}{\partial \gamma_{ij}} 
   = \frac{\sqrt{\rm det\, \gamma}}{2\sqrt{x}} n^l \,{}^3\! f_{li} \tilde{V}^i_k \left({}^3\! f^{kj} - n^k n^j \right)\, , \\
   \label{Ug+Uf}
   &\frac{\partial U}{ \partial \gamma_{ij}} \gamma_{jk} 
   + \frac{\partial U}{ \partial \,{}^3\! f_{ij}} \,{}^3\! f_{jk} 
   = - \frac{1}{2\sqrt{x}} n^i \tilde{V}^m_k \,{}^3\! f_{ml} n^l\, .
\end{align}
We also use the Poisson brackets between $R_0^{(g)}$ and $R_i^{(g)}$.
\begin{align}\label{R0R0}
   &M_g^2\{ R_0^{(g)} (x)\, , \, R_0^{(g)} (y) \} 
   = - \left[ R^i_{(g)} (x) \frac{\partial}{\partial x^i} \delta^3 (x-y) 
   - R^i_{(g)} (y) \frac{\partial}{\partial y^i} \delta^3 (x-y) \right]  \, ,\\
   \label{R0Ri}
   &M_g^2 \{ R_0^{(g)} (x)\, , \, R_i^{(g)} (y) \} 
   = - R^0_{(g)} (y) \frac{\partial}{\partial x^i} \delta^3 (x-y) \, , \\
   \label{RiRj}
   &M_g^2 \{ R_i^{(g)} (x)\, , \, R_j^{(g)} (y) \} 
   = - \left[ R_j^{(g)} (x) \frac{\partial}{\partial x^i} \delta^3 (x-y)
   - R_i^{(g)} (y) \frac{\partial}{\partial y^j} \delta^3 (x-y) \right] \, .
\end{align}
The same relations hold for $R_0^{(f)}$ and $R_i^{(f)}$ 
with $M_g^2$ replaced with $M_f^2$.

Next, we present the calculations of Poisson brackets between three primary constraints, 
$\{ C^L\, , \, C^L \}$, $ \{ C^L_i(x)\, , \, \mathcal{C}(y) \}$, 
$\{ C^L_i\, , \, C^L \}$, and $\{ \mathcal{C}\, ,\, C^L \}$. 
We do not show the other Poisson brackets, 
$\{ \mathcal{C}\, , \, \mathcal{C} \}$ and $\{C^L_i\, ,\, C^L_j \}$, 
because $\{ \mathcal{C}\, , \, \mathcal{C} \}$ has been already confirmed 
to be proportional to $\mathcal{C}$ with the aid of $\mathcal{C}_i =0$ 
in Ref.~\cite{Hassan:2011ea} and 
$\{C^L_i\, ,\, C^L_j \}$ is obviously proportional to 
the constraint $C^L_i$ itself by use of Eq.~\eqref{RiRj}. 
Notice that we do not use $C^L = 0$ to confirm 
$\{ \mathcal{C}\, , \, \mathcal{C} \} =0$ and $\{C^L_i\, ,\, C^L_j \}=0$.

\begin{itemize}
\item $\{ C^L\, , \, C^L \}$

Here, we should note that $U$ depends on $\pi^{ij}$ only through $n^k$.
Since $n^k$ can be treated as fixed in $C^L$ owing to $\mathcal{C}_i=0$, 
the contribution of $\partial \left(\sqrt{\rm det \, \gamma}\,U \right) / \partial \pi^{ij}$ vanishes.
Therefore the term $\{ R_0^{(g)}(x) \, ,\, \sqrt{\rm det\, \gamma }\, U (y) \}$, 
which is proportional to $\delta^3(x-y)$ because $R_0^{(g)}$ and $U$ do not contain
derivatives of $\pi^{ij}$ and $\gamma_{ij}$, is canceled by its conjugate 
$\{ \sqrt{\rm det\, \gamma }\, U (x) \, , \, R_0^{(g)}(y) \}$.
As a result, we have
\begin{align}
   \{ C^L (x)\, , \, C^L (y) \} = M_f^4  \{ R_0^{(f)} (x)\, , \, R_0^{(f)} (y) \} 
   + M_g^4 n^i n^j  \{ R_i^{(g)} (x)\, , \, R_j^{(g)} (y) \}  \nonumber \\
   - \left[ 2m^2 M_{\rm eff}^2\, n^i \frac{\partial R_i^{(g)} }{ \partial \pi^{mn}} 
   \frac{\partial \left( \sqrt{\rm det \, \gamma }\, U \right) }{ \partial \gamma_{mn}} 
   - ( x \leftrightarrow y) \right]\, .
\end{align}
To proceed the calculation, we introduce localized smoothing functions $F(x)$ and $G(x)$ 
and take inner products of the smoothing functions and $C^L$ as 
\begin{align}
   \left< F C^L \right> := \int d^3x\, F(x) C^L(x)\, , \ \ \ \left< G C^L \right> := \int d^3y\, G(y) C^L(y)\, .
\end{align}
Then we compute, 
\begin{align}
   \{ \left< F C^L \right>\,, \, \left< G C^L \right> \} 
   = \int d^3x \int d^3y\, F(x) G(y) \{ C^L(x) \, , \, C^L(y) \} \, ,
\end{align}
from which we can extract $\{ C^L(x)\, , \, C^L(y) \}$.
Using Eqs.~\eqref{R0R0} and \eqref{RiRj}, we find
\begin{align}
   \{ \left< F C^L \right>\,, \, \left< G C^L \right> \} 
   &= \left<  \left( - M_f^2 R^{n\, (f)} -M_g^2 n^n n^j R_j^{(g)} 
   - 4\tilde{m}^2 M_g^2 n^i \gamma_{im} \frac{\partial \left( \sqrt{\rm det \, \gamma }\, U \right) }
   { \partial \gamma_{mn}} \right) F \partial_n G - (F \leftrightarrow G)  \right>\,  \nonumber \\
   &= M_g^2 \left< \left( {}^3\! f^{ij} - n^i n^j \right) \left( R_j^{(g)}
   - 2\tilde{m}^2 \, \frac{\sqrt{\rm det \, \gamma }}{\sqrt{x}} \tilde{V}^k_j\, {}^3\! f_{kl} n^l \right) 
   (F\partial_i G) - (F \leftrightarrow G) \right>\, .
\end{align}
Here we have used the constraint $C^L_i =0$ and Eq.~\eqref{gnU} in the second equality.
Therefore we can confirm that under the constraint $\mathcal{C}_j=0$, 
$\{ C^L(x)\, , \, C^L(y) \}$ vanishes.
Note that we have not used $C^L=0$ here. 

\item  $\{ C^L_i\, , \, \mathcal{C} \} $

We calculate 
\begin{align}
   \{ C^L_i(x)\, , \, \mathcal{C}(y) \} 
   = M_g^4 \left[ \{ R_i^{(g)} (x) \, , \, R_0^{(g)} (y) \} 
   + \{ R_i^{(g)} (x) \, , \, R_k^{(g)} D^k_l n^l  (y) \} 
   + \{ R_i^{(f)} (x) \, , \, D^k_l n^l (y) \} R_k^{(g)} (y) \right. \nonumber \\
   \left. + 2\tilde{m}^2 \{ R_i^{(g)} (x) \, , \,  \sqrt{\rm det \, \gamma }\, V(y) \}
   + 2\tilde{m}^2 \{ R_i^{(f)} (x) \, , \,  \sqrt{\rm det \, \gamma }\, V(y) \} \right]\, ,
\end{align}
which should vanishes because $C^L_i$ is just the generator of the spatial translation and 
constraint $\mathcal{C}(y)$ is satisfied over the whole spatial hypersurface.
Introducing localized smoothing functions $F^i(x)$ and $G(y)$, we compute 
\begin{align}\label{FiCiGC}
   \{ \left< F^i C^L_i \right>\,, \, \left< G\, \mathcal{C} \right> \}
   &= M_g^2 \left< - \left(F^j \partial_j G\right) R_0^{(g)} - \left( F^j \nabla^{(g)}_j \left( G D^m_l n^l \right) 
   - \left( \nabla^{(g)}_i F^m \right) G D^i_l n^l \right) R_m^{(g)} \right. \nonumber \\
   &\ \ \ \left. + 2G \left( \nabla_i^{(g)} F^m \right) \gamma_{mj} 
   \left( R_k^{(g)} \frac{\partial \left( D^k_l n^l \right)}{\partial \gamma_{ij}} 
   + 2\tilde{m}^2 \frac{\partial \left( \sqrt{\rm det\, \gamma}\, V \right)}{\partial \gamma_{ij}} \right) 
   \right. \nonumber \\
   &\ \ \  \left.+ 2G \left( \nabla_i^{(f)} F^m \right) {}^3\! f_{mj} 
   \left( R_k^{(g)} \frac{\partial \left( D^k_l n^l \right)}{\partial {}^3\! f_{ij}} 
   + 2\tilde{m}^2 \frac{\partial \left( \sqrt{\rm det\, \gamma}\, V \right)}
   {\partial {}^3\! f_{ij}} \right) \right>\,  \nonumber \\
   &=M_g^2 \left< - \left(F^j \partial_j G\right) \left( R_0^{(g)} + D^l_kn^k R_l^{(g)} \right)
   - F^k G R_j^{(g)} \nabla_k^{(g)} \left( D^j_l n^l \right) \right. \nonumber \\
   &\ \ \ \left. + G \left( \nabla_i^{(g)} F^m \right) \left[ D^i_l n^l R_m^{(g)} 
   + 2\tilde{m}^2 \frac{ \sqrt{\rm det \, \gamma} }{\sqrt{x}} 
   \left( - \tilde{V}^n_m \left(D^{-1} \right){}^l_n \,{}^3\! f_{lj} \gamma^{ij} 
   + \sqrt{x}^2 \tilde{V}^l_m D^i_l \right) \right] \right.\nonumber \\
   &\ \ \ \left. + G \left( \nabla_m^{(g)} F^m \right) 2 \tilde{m}^2 \sqrt{\rm det \, \gamma} \, V
   - 2\tilde{m}^2 G\, \delta \Gamma^m{}_{ik}F^k  \sqrt{\rm det \, \gamma} \, 
   \sqrt{x}\, \tilde{V}^l_m D^i_l \right>\, .
\end{align}
In the second equality, we have used the relations \eqref{Vg}, \eqref{Vf} 
and the constraint $\mathcal{C}_k=0$, 
and rewrote $\nabla^{(f)}$ to $\nabla^{(g)}$ using the difference of the connections 
associated with the two metrics,
\begin{align}\label{dG}
   \delta \Gamma^m{}_{ik} := \Gamma_{(g)}^m{}_{ik} -  \Gamma_{(f)}^m{}_{ik}\, .
\end{align}
The quantity in the square brackets in the final expression of Eq.~\eqref{FiCiGC} vanishes 
using the relation
\begin{align}
   - \left(D^{-1}\right){}^l_m \,{}^3\! f_{lj} \gamma^{ij} + \sqrt{x}^2 D^i_m 
   = - {}^3\! f_{ml} n^l D^i_k n^k\, , 
\end{align}
which is derived from the definition of the matrix $D$ and $x$, 
with the aid of the constraint $\mathcal{C}_m=0$.
One can argue that the remainig terms also cancel with each other by the following discussion.
Integrating by parts the fourth term in the final expression of Eq.~\eqref{FiCiGC} 
produces the term in which $\nabla^{(g)}_m$ operates on $G$, 
which cancels the first term with the aid of the constraint $\mathcal{C}=0$, 
and the one in which $\nabla^{(g)}_m$ operates on $V$, which is rewritten as 
\begin{align}\label{nablaV}
   \nabla^{(g)}_m V &= \frac{\partial V}{\partial \left( \sqrt{x} D^a_b \right)} 
   \nabla^{(g)}_m \sqrt{x} D^a_b \, , \nonumber \\
   &= -\sqrt{x} \tilde{V}^i_j D^j_k \delta \Gamma^k{}_{im} 
   - \frac{1}{\sqrt{x}} \tilde{V}^i_j \,{}^3\! f_{ik}n^k \nabla^{(g)}_m \left( D^j_l n^l \right)\, , 
\end{align}
where we used 
\begin{align}
\nabla_a^{(g)} \,{}^3\! f_{ij} = \left( \nabla_a^{(g)} - \nabla_a^{(f)} \right) \,{}^3\! f_{ij}
= -2\,{}^3\! f_{l(i} \delta \Gamma^l_{j)a}\, .
\end{align}
Substituting Eq.~\eqref{nablaV} into Eq.~\eqref{FiCiGC}, 
the last term in Eq.~\eqref{FiCiGC} is canceled by the first term in Eq.~\eqref{nablaV}, and 
the second term in Eq.~\eqref{FiCiGC} is combined with the second term 
in Eq.~\eqref{nablaV} to be zero owing to the constraint $\mathcal{C}_j=0$.
Therefore all terms in Eq.~\eqref{FiCiGC} are canceled 
and $\{ C^L_i(x)\, , \, \mathcal{C}(y) \} =0$ is proved.
Here we have not used $C^L=0$ as in the case of $\{ C^L\, , \, C^L \}$.

\item $\{ C^L_i\, , \, C^L \}$

Here we also introduce localized smoothing functions $F^i(x)$ and $G(y)$ and compute 
\begin{align}\label{FiCiGCL}
   \{ \left< F^i C^L_i \right>\,, \, \left< G C^L \right> \}
   &= \left< -M_f^2 \left( F^k \nabla^{(f)}_k G \right) R_0^{(f)} 
   - M_g^2\left( F^k \nabla^{(g)}_k \left( Gn^j \right) 
   - \left( \nabla^{(g)}_k F^j \right) Gn^k \right) R_j^{(g)} \right. \nonumber \\
   & \left.\ \ \  + 4\tilde{m}^2M_g^2 \left[ G\left( \nabla^{(g)}_i F^m \right) \gamma_{mj} 
   \frac{\partial \left( \sqrt{\rm det\, \gamma} \,U \right)}{\partial \gamma_{ij}} 
   + G\left( \nabla^{(f)}_i F^m \right) \,{}^3\! f_{mj} \frac{\partial \left( \sqrt{\rm det\, \gamma} \,U \right)}
   {\partial \,{}^3\! f_{ij}}  \right] \right>\, , \nonumber \\
   &= 2\tilde{m}^2 \left< \left( F^k \nabla^{(f)}_k G \right) \sqrt{\rm det \, \gamma}\,U
   - \left\{ F^k \nabla^{(g)}_k n^j - \left( \nabla^{(g)}_k F^j \right)n^k \right\} G
   \frac{\sqrt{\rm det\, \gamma}}{\sqrt{x}} \tilde{V}^n_j\,{}^3\! f_{nl} n^l \right. \nonumber \\ 
   &\left. \ \ \ + G\left( \nabla^{(g)}_i F^m \right) \sqrt{\rm det\, \gamma}\, 
   \left[ \delta^i_m U - \frac{1}{\sqrt{x}} n^i \tilde{V}^l_m\, {}^3\! f_{lk}n^k \right]
   - 2G \delta \Gamma^m{}_{ik} F^k\,{}^3\! f_{mj} \frac{\partial U}{\partial\,{}^3\! f_{ij}}  \right> \, ,
\end{align}
which is also expected to vanish as in the case of $\{ C^L_i(x)\, , \, \mathcal{C}(y) \}$.
Here we have used the constraints $\mathcal{C}_j=0$ and $C^L=0$ 
and the relation \eqref{Ug+Uf} in the second equality.
We find that the second term in the braces and the second term in the square brackets 
cancel each other.
The first term can be integrated by parts as
\begin{align}\label{PI}
   &\left< \left( F^j \nabla_j^{(g)} G\right) \sqrt{\rm det \, \gamma}\, U \right> \nonumber \\
   &\ \ = -G \left( \nabla^{(g)}_j F^j \right) \sqrt{\rm det\, \gamma}\, U 
   -G F^j \left(\nabla^{(g)}_j n^i \right) \frac{\partial \left( \sqrt{\rm det\, \gamma} \,U \right)}{\partial n^i}
   - GF^j \left( \nabla^{(g)}_j {}^3\! f_{ab} \right) \frac{\partial \left( \sqrt{\rm det\, \gamma} \,U \right)}
   {\partial \,{}^3\! f_{ab}}\, \nonumber \\
  &\ \ = -G \left( \nabla^{(g)}_j F^j \right) \sqrt{\rm det\, \gamma}\, U 
  + G F^j \left(\nabla^{(g)}_j n^i \right) \frac{\sqrt{\rm det\, \gamma}}{\sqrt{x}} 
  \tilde{V}^k_i \,{}^3\! f_{kl}n^l + 2 \sqrt{\rm det\, \gamma} \,
  GF^j \,{}^3\! f_{la} \delta \Gamma^l{}_{bj} 
  \frac{\partial U}{\partial \,{}^3\! f_{ab}}\, ,
\end{align}
where we used the relation \eqref{Un} and 
the definition of the connection $\delta \Gamma$, \eqref{dG}.
From this equation \eqref{FiCiGCL}, 
we find that three terms in Eq.~\eqref{PI} are canceled by  
the first term in the square brackets, the terms proportional to 
$\tilde{V}$ and the last term in Eq.~\eqref{FiCiGCL}, respectively. 
Hence, the Poisson bracket between $C^L_i$ and $C^L$ 
is confirmed to vanish. 
Notice that we used $C^L=0$ here, unlike the previous calculations. 

\item $\{ \mathcal{C}\, ,\, C^L \}$

Finally we give an explicit expression for $\{ \mathcal{C}\, ,\, C^L \}$.
As before, we compute
\begin{align}
   &\{ \left< F \mathcal{C} \right>\,, \, \left< G C^L \right> \}  \nonumber \\
   &= M_g^2 \left< n^j R_j^{(g)} G \nabla_i^{(g)} \left( D^i_k n^k F \right)
   + 4\tilde{m}^2 \frac{ \partial \sqrt{{\rm det}\, \gamma}\, U}{ \partial \gamma_{mn}} 
   \gamma_{im}G \nabla_n^{(g)} \left( D^i_k n^k F \right)
    \right. \nonumber \\
   & \qquad + 2\tilde{m}^2 \frac{\sqrt{{\rm det}\, \gamma}}{ \sqrt{x}} 
   \left(D^{-1}\right)^k_l \tilde{V}^m_k \,{}^3\!f_{mj} \gamma^{jn} F \nabla^{(g)}_n \left( n^l G \right) 
   \nonumber \\
   &\left. \qquad 
   - \tilde{m}^2 \left[ \frac{2}{\sqrt{{\rm det}\, \gamma} }
   \frac{ \partial \sqrt{{\rm det}\, \gamma}\, U}{ \partial \gamma_{mn}}
   \left( \pi \gamma_{mn} -2\pi_{mn} \right) 
   + \frac{ \sqrt{{\rm det}\, \gamma}}{ \sqrt{{\rm det}\, {}^3\!f }} \sqrt{x} 
   \tilde{V}^m_l \,{}^3\!f^{lk} D^n_k \left( p \,{}^3\!f_{mn} -2p_{mn} \right)  \right] FG
    \right>\, .
\end{align}
Here we have used  the constraints $\mathcal{C}_i = 0$ and $\mathcal{C} = 0$.
The last term includes no derivative and is linear in $p^{mn}$, which appears 
only in the squared or differentiated form in all constraints.
Therefore this term cannot be removed 
and hence $\{ \mathcal{C}\, ,\, C^L \}$ never becomes zero.

\end{itemize}

Hence, we conclude that all the Poisson brackets among the primary constraints 
$\mathcal{C}$, $C^L$, and $C^L_i$ are zero, except for $\{ \mathcal{C}\, ,\, C^L \}$, 
as is expected from the calculation in Ref.~\cite{HR1}.
We stress that we used $C^L=0$ only to show $\{ C^L_i\, ,\, C^L \}=0$, 
which allows us to simplify the calculation in the presence of a doubly coupled matter, 
which will be discussed in Sec.~3.

\section{doubly coupled matter}

Next, we examine the existence of the BD ghost in the presence of a doubly coupled matter field.
The proof in Sec.~2 can be easily extended to the case in which there is matter 
that couples to either metric, but not to both. 
This is because the matter contribution can be totally absorbed 
by the redefinition of $R^{(g)}_0$ and $R^{(g)}_i$ (or $R^{(f)}_0$ and $R^{(f)}_i$) 
in all constraints and the fundamental Poisson brackets \eqref{R0R0}-\eqref{RiRj} 
hold without any change.

If we consider the case in which there is a matter field which is
coupled to both metrics, however, the proof will in general break
down. In this case, it is impossible to define such a new shift-like
vector which makes the Lagrangian linear in the both lapse
functions. Hence, we will not have the constraint that corresponds to
$\mathcal{C}$ in this case, and hence the model will possess the BD
ghost. One particular example of a ghost-free theory with a matter
field which couples to both metrics is
\begin{eqnarray}
S & = & \int
 d^{4}x\sqrt{-g}\left[\frac{M_{g}^{2}R^{(g)}}{2}+2m^2 M_{\rm{eff}}^2 
\sum_{n}c_ne_{n}\!\!\left(\sqrt{g^{\mu\nu}(f_{\mu\nu}+\alpha
		     \partial_{\mu}\phi\,\partial_{\nu}\phi)}\right)\right]
\cr
&  &\quad + 
\int d^{4}x\sqrt{-f}\left[\frac{M_{f}^{2}R^{(f)}}{2}-\frac{1}{2}\, f^{\mu\nu}\partial_{\mu}\phi\,\partial_{\nu}\phi\right]\, ,\label{int}
\end{eqnarray}
where $\alpha$ is a constant. This model is free from the BD ghost
because 
we can rewrite the action into the one with no coupling to the
metric $g_{\mu\nu}$, making a field redefinition as
\begin{equation}
f_{\mu\nu}\to\tilde{f}_{\mu\nu}-\alpha
 \partial_{\mu}\phi\,\partial_{\nu}\phi\,.
\label{eq:replace}
\end{equation}
In the remaining kinetic term, higher temporal
derivative terms do not arise after the replacement \eqref{eq:replace},
for the same reason why the temporal derivatives of the lapse and
shift can be removed by integrating by parts in the Einstein-Hilbert action.
This possibility seems the natural and ghost-free extension of the
model proposed in \cite{DeFelice:2013tsa} (see also \cite{MGonefield}) to the bigravity case. 

\subsection{Seeking for models with doubly coupled matter which have no ghost}

We have introduced some models that have no BD ghost. In this section we
want to show that, as soon as we further extend the proposed ghost-free
action, the ghost will appear. To prove the existence of the ghost,
it will suffice to give a proof of its existence on some reasonable
background.

Let us analyze then the Lagrangian 
\begin{equation}
\mathcal{L}\ni\sqrt{-g}\, P(X,\phi)+\sqrt{-f}\,\tilde{P}(\tilde{X},\phi)\,,
\end{equation}
where with $P$ and $\tilde{P}$ are arbitrary functions of $\{\phi,\, X\}$
and $\{\phi,\,\tilde{X}\}$, respectively, with
\begin{equation}
X=-\frac{1}{2}\, g^{\alpha\beta}\partial_{\alpha}\phi\partial_{\beta}\phi\,,\qquad\qquad\tilde{X}=-\frac{1}{2}\, f^{\alpha\beta}\partial_{\alpha}\phi\partial_{\beta}\phi\,.
\end{equation}

\subsubsection{The Friedmann background}

As a first step, let us study scalar-type linear perturbations
around the FLRW background. We fix the gauge so that the metrics are
given by 
\begin{eqnarray*}
ds^{2} & = & -a^{2}(1+2\alpha)\, dt^{2}-2\partial_{i}\chi\, dtdx^{i}+a^{2}\delta_{ij}dx^{i}dx^{j}\,,\\
d\tilde{s}^{2} & = & -\tilde{c}^{2}\tilde{a}^{2}(1+2\tilde{\alpha})\, dt^{2}-2\partial_{i}\tilde{\chi}\, dtdx^{i}+\tilde{a}^{2}[\delta_{ij}(1+2\tilde{\Phi})+2\partial_{i}\partial_{j}\tilde{\gamma}]\, dx^{i}dx^{j}\,,
\end{eqnarray*}
and the scalar field by 
\begin{equation}
\phi=\bar{\phi}+\delta\phi\,.
\end{equation}
Therefore, in total, we have seven scalar variables describing the three-dimensional
scalar-sector of the perturbation fields. Following the standard way
of reducing the action for the perturbation variables, we first eliminate
the lapse and shift perturbations $\alpha$, $\chi$, $\tilde{\alpha}$,
and $\tilde{\chi}$, using the Hamiltonian and momentum constraints. 
If the BD ghost is present, we will be left with three scalar
variables with none of them being a Lagrange multiplier. In other words,
we can see whether the BD ghost is present or not, by evaluating the
determinant of the kinetic matrix $\mathcal{A}_{ij}$ (after integrating
out the auxiliary fields $\alpha$, $\chi$, $\tilde{\alpha}$, and
$\tilde{\chi}$).

We will find that this determinant does not vanish in general. If
it does not vanish, we can conclude that the BD ghost is present. However,
there might be a subclass of theories for which the ghost is absent:
at least we should see the disappearance of the ghost in the case
$\tilde{P}(\tilde{X},\phi)=\tilde{P}(\phi)$.

After some algebra, we find 
\begin{equation}
\det(\mathcal{A}_{ij})\propto \dot{a}^{2} \dot{\bar{\phi}}^{2} 
(6\,{\xi}^{2}c_{{3}}+4\,\xi\, c_{{2}}+c_{{1}})
\left(P_{,X}+2XP_{,{\it XX}}\right)
\left(\tilde{P}_{,\tilde{X}}+2\tilde{X}\tilde{P}_{,\tilde{X}\tilde{X}}\right) \,,
\end{equation}
where $\xi:=\tilde{a}/a$. First
of all, $\det(\mathcal{A}_{ij})$ vanishes when the background solution
satisfies $\Gamma\equiv c_{{1}}+4\,\xi\, c_{{2}}+6\,{\xi}^{2}c_{{3}}=0$
\footnote{We will not discuss here this solution, for two reasons. First, it
is a constraint on the choice of the background $\xi$, 
rather than a
condition that is identically satisfied by a given model. 
This implies that we cannot be sure
whether or not the BD ghost would be absent
on different backgrounds even for the same model.  
Second, this choice of the background $\xi$ is known to be pathological
in the sense that many modes (not only the BD scalar) are
absent in the quadratic action, and hence we do not consider this
case in the following discussion.%
}, or for the following sub-classes of theories 
\begin{equation}\label{FLRWP}
P_{,X}+2XP_{,{\it XX}}=0\,,\qquad\mathrm{or}\qquad\tilde{P}_{,\tilde{X}}+2\tilde{X}\tilde{P}_{,\tilde{X}\tilde{X}}=0\,.
\end{equation}
One solution 
 for Eq.~\eqref{FLRWP} 
is provided by the case $\tilde{P}(\tilde{X},\phi)=\tilde{P}(\phi)$.
However, we realize that there could also be another possibility,
that is 
\begin{equation}
P=P(X,\phi)\,,\qquad\tilde{P}=F_{1}(\phi)\sqrt{\tilde{X}}+F_{2}(\phi)\,,\qquad\mathrm{with}\qquad
 F_{1}\neq0\,.
\label{eq:BDfull}
\end{equation}
Notice that the roles of $P$ and $\tilde{P}$ can be exchanged.

This result implies that on the FLRW background, at least, at linear
level in the perturbation fields, the BD ghost is absent not only
for the known case $\tilde{P}(\tilde{X},\phi)=\tilde{P}(\phi)$, but
also for the more general class of models defined in (\ref{eq:BDfull}).
This result, at first sight, seems to contradict our previous calculations.
However, even if the BD ghost is absent on the FLRW background (at
linear level in the perturbation fields), we still need to check if
the BD ghost remains absent for this new class of models (\ref{eq:BDfull})
as we change the background. In fact, in the succeeding section
we will discuss perturbation in an anisotropic background for the models (\ref{eq:BDfull})
as a counterexample, for which the BD ghost is present.

\subsubsection{The anisotropic background}

As we have already seen, the study on the FLRW background has already
sufficiently restricted the possibilities for the choice of function
$\tilde{P}$ to the case $\tilde{P}=F_{1}(\phi)\sqrt{\tilde{X}}+F_{2}(\phi)$.
In order to understand whether this model does possess the BD ghost
(or not) on other general backgrounds, we found it convenient to focus
on the simplest example of an anisotropic background.

Focusing on a particular anisotropic Bianchi-I type manifold with residual axial symmetry, 
we will
consider a sub-case where we have $F_{1}=\mathrm{constant}$ and $F_{2}=0$
in (\ref{eq:BDfull}). The field profile compatible with the Bianchi
I symmetry can be written then as 
\begin{equation}
\phi=v_{0}\int^{t}\frac{\tilde{c}(\eta)}{\tilde{f}_{1}(\eta)}\, d\eta+v_{1}x+\delta\phi\,,
\end{equation}
where we have chosen the time-profile for the field $\phi$, leaving
$g_{00}=-N(t)^{2}$ as a free variable to be determined by the equations
of motion. We have also introduced a possible non-trivial $x$-profile
$v_{1}x$ for the field, which is still consistent with the symmetry
of the background, as we have imposed $\partial\tilde{P}/\partial\phi=0$.
Indeed this background-profile for $\phi$ implies that 
\begin{equation}
\tilde{X}=\frac{4v_{1}^{2}}{\tilde{f}_{1}^{2}}\,,
\end{equation}
where, for simplicity, we have also fixed $v_{0}=3v_{1}$. Finally
we further restrict our attention to the case with the simplest linear
functions $P(X)$, so that, in total, the contribution of this scalar
field to the Lagrangian reads 
\begin{equation}
\mathcal{L}\ni\sqrt{-g}\, C_{1}X+\sqrt{-\tilde{g}}\tilde{C}_{1}\sqrt{\tilde{X}}\,,
\end{equation}
where $C_{1}$ and $\tilde{C}_{1}$ are numerical constants. If the
BD ghost is present for this simple example, it will also be present
for the general class of theories defined in (\ref{eq:BDfull}).

For ghost-free models on this axial-symmetric Bianchi-I background, 
 we should have in
total only 3 odd (vectors under 2d rotations) modes and 5 even (scalars
under 2d rotations) modes. We consider the even modes. With an appropriate
choice of gauge, the two metrics can be written as 
\begin{eqnarray}
ds^{2} & = & -N^{2}(1+\alpha_{1})dt^{2}+2\partial_{x}\chi_{1}\, dtdx+2\partial_{i}\eta_{1}dtdy^{i}+f_{1}^{2}(1+\beta_{1})dx^{2}+f_{2}^{2}\delta_{ij}dy^{i}dy^{j}\,,\\
d\tilde{s}^{2} & = & -\tilde{c}^{2}(1+\alpha_{2})dt^{2}+2\partial_{x}\chi_{2}\, dtdx+2\partial_{i}\eta_{2}dtdy^{i}+2(\partial_{x}\partial_{i}\zeta_{2})\, dxdy^{i}\nonumber \\
 &  & {}+\tilde{f}_{1}^{2}(1+\beta_{2})dx^{2}+\tilde{f}_{2}^{2}[\delta(1+\Phi_{2})+\partial_{i}\partial_{j}\gamma_{2}]\, dy^{i}dy^{j}\,,
\end{eqnarray}
where $\{y^1,y^2\} \equiv \{y,z\}$.

Next, in order to reduce the action, we need to integrate out all
the known auxiliary fields, which are the six lapse and shifts, $\alpha_{1},\chi_{1},\eta_{1},\alpha_{2},\chi_{2},\eta_{2}$. 
After integrating out the auxiliary fields in the even modes, we are
left with $12-6=6$ even modes (whereas, as for the odd sector, as
already mentioned, the gauge fixing removes one mode and the momentum
constraint other two, so that only 3 odd modes remain). In total (for
the even and odd sectors) we are left with nine propagating modes.
However, if the BD ghost were absent, we would find that the determinant
of the kinetic matrix for the even modes should vanish, i.e.\ one
of the six even modes should reduce to a Lagrange multiplier.

To simplify the analysis we focus on a background which is an exact
solution of the equations of motion, defined by 
\begin{eqnarray}
\tilde{f}_{2}(t) & = & (\xi+w\bar{\xi}_{2})\, f_{2}(t)\,,\\
\tilde{f}_{1}(t) & = & \xi\, f_{1}(t)\,,\\
\tilde{c}(t) & = & (\xi+w\bar{\xi}_{0})\,,\\
N(t) & = & 1\,,\\
f_{1}(t) & = & 1\,,\\
f_{2}(t) & = & e^{wt}\,,
\end{eqnarray}
where $\xi$, $\bar{\xi}_{0}$, $\bar{\xi}_{2}$, and $w$ are all constants. These
constants, together with the parameters of the theory $c_{0}$, $c_{1}$, $c_2$, $c_{3}$, 
$c_{4}$, $C_{1}$, and $\tilde{C}_{1}$are constrained to solve the six independent
equations of motion. We solve the six independent equations of motion
for the variables $\tilde{C}_{1}$, $c_{0}$, $c_{1}$, $c_{3}$,
$c_{4}$, and $C_{1}$ in terms of $v_{1}$, $\xi$, $\bar{\xi}_{0}$,
$\bar{\xi}_{2}$, $w$ and $\kappa:=M_{f}^{2}/M_{g}^{2}$. Notice that the background has
a Minkowski limit, that is $w\to0$, in which the quantity 
\begin{equation}
\Gamma = c_{1}+4c_{2}\xi+6c_{3}\xi^{3}\approx\frac{2\xi M_{g}^{2}(1+\kappa\xi^{2})(4\bar{\xi}_{0}+5\bar{\xi}_{2})}{\bar{\xi}_{0}^{2}(\bar{\xi}_{0}-\bar{\xi}_{2})m^{2}M_{\rm eff}^2}+\mathcal{O}(w)
\end{equation}
does not necessarily vanish. This shows that this Minkowski limit
does not, in general, belong to the unhealthy branch of the FLRW background,
defined by the condition $\Gamma=0$.

Although it is not necessary in this context, we will fix the parameters
imposing 
\begin{equation}
(4\bar{\xi}_{0}+5\bar{\xi}_{2})(\bar{\xi}_{0}-\bar{\xi}_{2})>0\,,
\end{equation}
(to have a positive $\Gamma$ for the stability of the background).
In this Minkowski limit, using the constraint equations, we also find
\begin{eqnarray}
C_{1} & \approx & \frac{\xi^{3}}{4v_{1}^{2}}\,|\tilde{C}_{1}v_{1}|\,,\\
|\tilde{C}_{1}v_{1}| & \approx & \frac{4M_{g}^{2}(1+\kappa\xi^{2})}{\xi^{2}\bar{\xi}_{0}}\, w\,,
\end{eqnarray}
so that $w/\bar{\xi}_{0}>0$.

As already mentioned earlier, in order to see whether the BD ghost
is absent or not, we need to study the six by six kinetic matrix for
the even modes which remain after having integrated out the auxiliary
variables. We study the eigenvalues of such a kinetic matrix in the
large momenta limit (i.e.\ for large $k$ and $q$), where all variables
are supposed to be expanded by the Fourier mode $e^{-i(kx+q_{1}y+q_{2}z)}$
with $q=\sqrt{q_{1}^{2}+q_{2}^{2}}$.

We find that the kinetic matrix $\mathbf{K}$ has in general a non-zero determinant, i.e.\ 
the BD ghost will be present in general. To make it evident, we expand
the determinant (around $w\to0$) to find 
\begin{equation}
\det(\mathbf{K})=-\frac{5}{4096}\,\frac{k^{2}q^{8}M_{g}^{12}(1+\kappa\xi^{2})^{2}\kappa^{3}\xi^{5}w}{4v_{1}^{2}(2k^{2}+q^{2})^{2}\bar{\xi}_{0}^{3}}<0\,.
\end{equation}
To be more precise, we have studied the sign of all six eigenvalues
  and checked that only one of them is negative in this limit, which makes the determinant negative. 
Therefore, on axial symmetric
anisotropic Bianchi-I manifolds, we can conclude that the case
$\tilde{P}\propto\sqrt{\tilde{X}}$ has 1) one more degree of freedom
than models which are free from the BD ghost, and 2) the extra mode is
indeed a ghost, at least on the Minkowski limit of such an anisotropic
solution.

Finally, from the analysis of perturbations around the FLRW and anisotropic
backgrounds, we conclude that only $\tilde{P}=\tilde{P}(\phi)$ is
the possibility to avoid the BD ghost in the class of models defined
by (\ref{eq:BDfull}).

\section{proof of the absence of ghost in the case $\tilde{P} = \tilde{P} (\phi)$}
 
In this section we give a proof of the absence of ghost in the case $\tilde{P} = \tilde{P} (\phi)$.
We consider a k-essence scalar field $\phi$ which couples to both metrics, 
but whose kinetic term contains only the physical metric $g_{\mu\nu}$.
Namely, the Lagrangian of the scalar field is 
\begin{align}\label{Lm}
 \mathcal{L}_{m} = \sqrt{-g}\, P(X,\, \phi) + \sqrt{-f}\, \tilde{P}(\phi) \, .
\end{align}
The case in which the kinetic term of matter contains only the hidden metric 
can be also treated in the same way.
The conjugate momentum of $\phi$ is derived from this action as 
\begin{align}
   \pi_{\phi} = N^{-1} \sqrt{{\rm det}\, \gamma}\, \left( \frac{\partial P}{\partial X} \right)_{\!\! \phi} 
   \left( \partial_t \phi - N^i \partial_i \phi \right)\, ,
\end{align}
where the subscript $\phi$ indicates that the partial differentiation is taken with $\phi$ fixed.
Using the definition of $X$, $\pi_{\phi}^2$ is written as
\begin{align} \label{pi2}
  \pi^2_{\phi} = {\rm det} \gamma \left( \frac{\partial P}{\partial X} \right)_{\!\! \phi}^2 
  \left[ 2X + \gamma^{ij} \partial_i \phi \partial_j \phi \right]\, .
\end{align}
This implies that $X$ can be expressed by $\gamma^{ij}$, $\phi$ and $\pi_{\phi}$ 
without using $N$, $N^i$ and $\partial_t \phi$.
Therefore we find that the matter Hamiltonian becomes
\begin{align}
   \mathcal{H}_m &= NC_m
   + N^i \pi_{\phi} \partial_i \phi + L \sqrt{{\rm det}\, {}^3\! f }\, \tilde{P}\, , 
\end{align}
where
\begin{align}\label{Cm1}
   C_m &:= \sqrt{{\rm det}\, \gamma}\, \left[ \left( \frac{\partial P}{\partial X} \right)_{\!\! \phi} 
   \left( 2X + \gamma^{ij} \partial_i \phi \partial_j \phi  \right) - P \right] \, .
\end{align}
This matter Hamiltonian $\mathcal{H}_m$ is linear in $N$, $N^i$ and $L$.
$C_m$ depends on $\gamma^{ij}$, $\phi$ and $\pi_{\phi}$, 
but $\pi_{\phi}$ appears in $C_m$ only through $X$, 
i.e. $C_m = C_m \left( \gamma^{ij},\, \phi,\, X\left( \gamma^{ij},\, \phi,\, \pi_{\phi} \right) \right)$.
After rewriting the shift vector $N^i$ using Eq.~\eqref{n},  the total Lagrangian becomes
\begin{align}\label{Ltot}
   \mathcal{L}_{total} =&  \pi^{ij} \partial_t \gamma_{ij} + p^{ij} \partial_t {}^3\! f_{ij} 
   + \pi_{\phi} \partial_t \phi \nonumber \\
   & +L^i \left[ M_g^2 R_i^{(g)} - \pi_{\phi} \partial_i \phi + M_f^2 R_i^{(f)} \right] \nonumber \\
   & + L \left[ M_f^2 R^{0(f)} + n^i \left( M_g^2 R_i^{(g)} - \pi_{\phi} \partial_i \phi \right) 
   + 2m^2 M_{\rm{eff}}^2 \sqrt{{\rm det}\, \gamma}\, U - \sqrt{{\rm det}\, {}^3\! f }\, \tilde{P} \right] \nonumber \\
   & +N \left[ M_g^2 R^{0(g)} -C_m 
   + D^i_j n^j \left( M_g^2 R_i^{(g)}  - \pi_{\phi} \partial_i \phi \right) 
    +  2m^2 M_{\rm{eff}}^2 \sqrt{{\rm det}\, \gamma}\, V \right]\, .
\end{align}
We redefine the constraints $\mathcal{C}$, $C^L$ and $C^L_i$ 
as the coefficient of $N$, $L$ and $L^i$ in Eq.~\eqref{Ltot}, respectively.
We also redefine the constraint $\mathcal{C}_i$ to the one 
obtained from the variation of $\mathcal{L}_{total}$ with respect to $n^i$ as 
\begin{align}
   \mathcal{C}_i := M_g^2 R_i^{(g)} - \pi_{\phi} \partial_i \phi
   - 2\tilde{m}^2 M_g^2 \sqrt{ {\rm det}\, \gamma}\, \frac{n^l \,{}^3\! f_{lj}}{\sqrt{x}} \tilde{V}^j_i\, .
\end{align}
Let us confirm that all the Poisson brackets of three constraints $\mathcal{C}$, $C^L$ and $C^L_i$ 
vanish except $\{ \mathcal{C},\, C^L \}$.

We introduce  
$ R_0^{(g) \prime}  := R_0^{(g)} - M_g^{-2}C_m $ and 
$ R_i^{(g) \prime} := R_i^{(g)} - M_g^{-2}\pi_{\phi} \partial_i\phi $, 
then the constraints $\mathcal{C}$, $\mathcal{C}_i$, $C^L_i$ takes the same forms as 
\eqref{C}, \eqref{Ci}, \eqref{CLi} with $R_0^{(g)}$ and $R_i^{(g)}$ 
replaced with $R_0^{(g) \prime}$ and $R_i^{(g) \prime}$.
We find that $ R_0^{(g) \prime}$ and $ R_i^{(g) \prime}$ satisfy the same relations as 
Eqs.~\eqref{R0R0}, \eqref{R0Ri} and \eqref{RiRj}, as is shown below.

In order to show these relations, we have to evaluate the partial differentiations of $C_m$ 
with respect to $\phi$, $\pi_{\phi}$, and $\gamma_{mn}$.
First, we have
\begin{align}
   \left( \frac{\partial C_m}{\partial X} \right)_{\!\! \phi} = \sqrt{{\rm det}\, \gamma}\, A\, ,
\end{align}
where
\begin{align}
   A := \left( \frac{\partial^2 P}{\partial X^2} \right)_{\!\! \phi} 
   \left( 2X + \gamma^{ij} \partial_i \phi \partial_j \phi \right) 
   + \left( \frac{\partial P}{\partial X} \right)_{\!\! \phi}\, .
\end{align}
Using Eq.~\eqref{pi2}, we find 
\begin{align}
  \left( \frac{\partial X}{\partial \phi} \right)_{\!\! \pi_{\phi},\, \gamma^{ij}} 
  &= - A^{-1} \left[ \left( \frac{\partial P}{\partial X} \right)_{\! \phi} 
  \partial^i \phi \frac{\partial \left( \partial_i \phi \right)}{\partial \phi}
  + \frac{\partial^2 P}{\partial X \partial\phi} 
  \left( 2X + \gamma^{ij} \partial_i \phi \partial_j \phi \right) \right]  \, , \\
  \left( \frac{\partial X}{\partial \pi_{\phi}} \right)_{\!\! \phi,\, \gamma^{ij}} 
  &= \left( {\rm det} \gamma\, A \left( \frac{\partial P}{\partial X} \right)_{\! \phi} \right)^{-1} \pi_{\phi} \, .
\end{align}
Solving Eq.~\eqref{pi2} for $X$,  we find $X$ depends on $\gamma_{ij}$ only through the form 
$\frac{\pi_{\phi} }{ \sqrt{{\rm det}\, \gamma} }$ and $\gamma^{ij} \partial_i \phi \partial_j \phi$. 
Therefore the partial differentiations of $X$ with respect to $\gamma_{mn}$ is 
\begin{align}
   \left( \frac{\partial X}{\partial \gamma_{mn}} \right)_{\!\! \phi,\, \pi_{\phi}}
   = \frac{1}{2} A^{-1} \left[ \left( \frac{\partial P}{\partial X} \right)_{\!\! \phi} 
   \partial^m\phi \partial^n \phi - \frac{\pi_{\phi}^2}{ {\rm det} \, \gamma} 
   \left( \frac{\partial P}{\partial X} \right)_{\!\! \phi}^{-1} \gamma^{mn} \right]\, .
\end{align}
Using the above relations, we obtain the partial differentiations of $C_m$ with respect to $\phi$ as
\begin{align}\label{Cmphi}
   \left( \frac{\partial C_m}{\partial \phi} \right)_{\!\! \pi_{\phi},\, \gamma^{ij}} 
   &= \left( \frac{\partial C_m}{\partial X} \right)_{\!\! \phi,\, \gamma^{ij}} 
   \left( \frac{\partial X}{\partial \phi} \right)_{\!\! \pi_{\phi},\, \gamma^{ij}}
   +  \sqrt{{\rm det}\, \gamma}\, \left[ 2\left( \frac{\partial P}{\partial X} \right)_{\!\! \phi} \partial^i \phi 
   \frac{\partial \left( \partial_i \phi \right)}{\partial \phi}
   + \frac{\partial^2 P}{\partial X \partial\phi} \left( 2X + \gamma^{ij} \partial_i \phi \partial_j \phi \right)
   - \left( \frac{\partial P}{\partial \phi} \right)_{\!\! X} \right]\, , \nonumber \\
   &= \sqrt{{\rm det}\, \gamma}\, \left[ \left( \frac{\partial P}{\partial X} \right)_{\!\! \phi} \partial^i \phi 
   \frac{\partial \left( \partial_i \phi \right)}{\partial \phi}
   - \left( \frac{\partial P}{\partial \phi} \right)_{\!\! X} \right]  \, , 
\end{align}
and the one with respect to $\pi_{\phi}$ as
\begin{align}\label{Cmpi}
   \left( \frac{\partial C_m}{\partial \pi_{\phi}} \right)_{\!\! \phi,\, \gamma^{ij}} = \sqrt{{\rm det}\, \gamma}\, 
   \left( \frac{\partial P}{\partial X} \right)_{\!\! \phi}^{-1} \pi_{\phi} \, .
\end{align}
The partial differentiations of $C_m$ with respect to $\gamma_{mn}$ is written as 
\begin{align}\label{Cmg}
    \left(\frac{\partial C_m}{\partial \gamma_{mn}} \right)_{\! \phi,\, \pi_{\phi}}
   &= \frac{C_m}{2} \gamma^{mn} 
   - \sqrt{{\rm det}\gamma} \left( \frac{\partial P}{\partial X} \right)_{\! \phi} \partial^m \phi \partial^n \phi
   + \left( \frac{\partial C_m}{\partial X} \right)_{\! \phi}  
   \left( \frac{\partial X}{\partial \gamma_{mn}} \right)_{\phi,\, \pi_{\phi}}
   \, , \nonumber \\
   &= - \left( \frac{\sqrt{{\rm det}\, \gamma}}{2} P \gamma^{mn} 
   + \left( \frac{\partial P}{\partial X} \right)_{\! \phi} 
   \partial^m \phi \partial^n \phi \right) \, .
\end{align}

Now, we confirm that the same relations as Eqs.~\eqref{R0Ri} and \eqref{RiRj} 
hold for $ R_0^{(g) \prime}$ and $ R_i^{(g) \prime}$. 
Using Eqs.~\eqref{Cmphi} and \eqref{Cmpi}, 
the Poisson bracket between $ R_0^{(g) \prime}$ and itself becomes 
\begin{align}
   \left\{ \left<F M_g^2 R_0^{(g)\prime}  \right>,\, 
   \left<G  M_g^2 R_0^{(g)\prime}  \right> \right\} 
   = \left< \left[ M_g^2 R_{(g)}^{i\, \prime}  
   \left(G \partial_i F \right) \right] - \left[ F \leftrightarrow G\right] \right> \, .
\end{align}
Here, among the contribution from $C_m$, only the term containing 
$\frac{\partial \left( \partial_i \phi \right)}{\partial \phi}$ in Eq.~\eqref{Cmphi} 
escapes from the cancelation with its conjugate. 

Using Eq.~\eqref{Cmphi}, \eqref{Cmpi}, \eqref{Cmg}, we obtain 
the Poisson bracket between $ R_0^{(g) \prime}$ and $ R_i^{(g) \prime}$ as
\begin{align}
   &\left\{ \left<F^i  M_g^2 R_i^{(g)\prime}  \right>,\, 
   \left<G  M_g^2 R_0^{(g)\prime}  \right> \right\}  \nonumber \\
   &=  \left< -M_g^2 R_0^{(g)} F^i \partial_i G 
   - 2\gamma_{mi}  \frac{\partial C_m}{\partial \gamma_{mn}} G\nabla^{(g)}_n F^i
   + C_m F^i \partial_i G - \sqrt{{\rm det} \gamma}\, P G \nabla^{(g)}_i F^i 
   -\sqrt{{\rm det} \gamma}\, \partial_i \phi \partial^n \phi  \left( \frac{\partial P}{\partial X} \right)_{\phi} 
   G \nabla_n^{(g)} F^i \right> \, , \nonumber \\
   &= - \left<  M_g^2 R_0^{(g)\prime}  F^i \partial_i G \right> \, .
\end{align}
$\{ R_i^{(g) \prime} \, , \, R_j^{(g) \prime} \}$ is easily found to satisfy 
\begin{align}
M_g^2 \{ R_i^{(g) \prime} (x)\, , \, R_j^{(g) \prime} (y) \} 
   = - \left[ R_j^{(g) \prime} (x) \frac{\partial}{\partial x^i} \delta^3 (x-y)
   - R_i^{(g) \prime} (y) \frac{\partial}{\partial y^j} \delta^3 (x-y) \right] \, .
\end{align}
From the above calculations, we find that the same relations as 
Eqs.~\eqref{R0R0}, \eqref{R0Ri} and \eqref{RiRj} hold 
for $ R_0^{(g) \prime}$ and $ R_i^{(g) \prime}$. 
Recall that $C^L$ was not used to discuss 
the Poisson brackets among the constraints $\mathcal{C}$ and $C^L_i$. 
Hence, the calculations about $\mathcal{C}$ and $C^L_i$ in Sec.~2 do not change 
under the replacements $ R_0^{(g)} \rightarrow R_0^{(g) \prime}$ 
and $R_i^{(g)} \rightarrow R_i^{(g) \prime}$.
By contrast, when doubly coupled matter is present, $C^L$ becomes 
different from the one in Sec.~2 
by the term $\sqrt{{\rm det}\,{}^3\! f}\, \tilde{P}$ even after these replacements. 
This prohibits us simply applying the discussion in Sec.~2 
to the Poisson brackets related to $C^L$. \\

Now, let us look at the calculations of the Poisson brackets between 
$ \mathcal{C}$, $C^L$, $C^L_i$, one by one.

\begin{itemize}

\item $\{ \mathcal{C}\, , \, \mathcal{C} \}$, $\{ C^L_i\, , \, \mathcal{C} \} $ 
and $\{ C^L_i\, , \, C^L_j \} $ 

In the calculation of these Poisson brackets in Sec.~2, we did not use the constraints $C^L=0$, 
hence these brackets are proved to be zero just by replacing  
$ R_0^{(g)}$ and $R_i^{(g)}$ with $R_0^{(g) \prime}$ and $R_i^{(g) \prime}$, respectively.

\item $\{ C^L\, , \, C^L \} $

We can easily find that all the contribution of 
$\sqrt{{\rm det}\,{}^3\! f}\, \tilde{P}(\phi)$ cancels with its conjugate, 
because $\sqrt{{\rm det}\,{}^3\! f}\, \tilde{P}(\phi)$ does not depend 
on the derivatives of the variables. 
Therefore $\sqrt{{\rm det}\,{}^3\! f} \tilde{P}$ does not affect the calculation of $\{ C^L,\, C^L \}$ 
and we obtain $\{ C^L,\, C^L \}=0$ in the same way as in Sec.~2, using the replacements 
$ R_0^{(g)} \rightarrow R_0^{(g) \prime}$ and $R_i^{(g)} \rightarrow R_i^{(g) \prime}$.

\item $\{ C^L_i\, , \, C^L \} $

Focusing on $\sqrt{{\rm det}\,{}^3\! f}\, \tilde{P}(\phi)$ in $C^L$, we calculate
\begin{align}
   &\left\{ \left<F^i \left( M_f^2 R_i^{(f)} + M_g^2 R_i^{(g)\prime} \right) \right>,\, 
   \left<G  \sqrt{{\rm det}\,{}^3\! f} \tilde{P} (\phi) \right> \right\} \nonumber \\
   &\qquad\qquad\qquad 
   =\left<  \sqrt{{\rm det}\,{}^3\! f} \tilde{P} G \nabla^{(f)}_i F^i 
   + \partial_i \phi \sqrt{{\rm det}\,{}^3\! f} \frac{ \partial \tilde{P}}{\partial \phi} F^i G \right> \, 
   \nonumber \\
   &\qquad\qquad\qquad 
   = -\left< \sqrt{{\rm det}\,{}^3\! f} \tilde{P} F^i \partial_i G \right> \, .
\end{align}
This compensates the modification in $C^L=0$ used in the second equality of Eq.~\eqref{FiCiGCL}. 
The rest of calculations can be done in the same way as in Sec.~2 just by replacing 
$ R_0^{(g)}$ and $R_i^{(g)}$ with $R_0^{(g) \prime}$ and $R_i^{(g) \prime}$, respectively, 
to obtain $\{ C^L_i\, , \, C^L \}=0 $. 

\end{itemize}

From the above calculations, we have proven that the bigravity model has no BD ghost 
even in the presence of a doubly-coupled field matter whose Lagrangian is given by Eq.~\eqref{Lm}.

\section{Summary}

In this paper, we have shown that doubly coupled matter generally
brings the BD ghost. As a first step, we have presented the detailed 
proof of ghost-freeness using the Hamiltonian analysis in the bigravity model 
obtained as an extension of dRGT massive gravity.
The proof has been already given in Refs.~\cite{HR1,HR2,Hassan:2011ea}, 
but we have explicitly shown that four Lagrange multipliers remain unspecified 
after all the Poisson brackets among the constraints close, 
which is requested by the general covariance, using Hamiltonian formulation.

Next, we extended the above proof to the case with matter.
The extension of the proof of ghost-freeness to the case with matter 
can follow in the same way as in GR as long as matter only
couples to either metric. However, we found that the BD ghost
is present when matter couples to both metrics through the kinetic terms of the matter.
To prove the appearance of the BD ghost, we have considered linear perturbations
on the Friedmann and Bianchi-I backgrounds. 
We found that there really exists an extra degree of freedom 
and, at least, one mode becomes ghost 
in the Minkowski limit of an anisotropic solution. Furthermore, we
show that the BD ghost is absent when either of the two metrics is 
coupled only through the potential term using Hamiltonian analysis.

Therefore, the form of doubly coupled matter is considerably restricted 
not only by experiments of the equivalence principle, 
but also by the condition to avoid the BD ghost,. 
The result seems inconsistent with the idea to embed the ghost-free bigravity 
into compactified higher dimensional models~\cite{deRham:2013awa, Yamashita:2014cra}. 
As long as we consider the healthy branch of bigravity, we cannot
trace back the evolution of the universe beyond the energy scale determined
by the bare graviton mass $m$ \cite{Comelli:2012db, DeFelice:2014nja, Konnig:2014xva}. 
Hence, in order to take a ghost-free
bigravity model as a realistic model of our universe, we need to find
a way to derive such a model at low energies starting with a more
complete model which is valid even at high energies. One possibile
completion might be obtained by considering braneworld models. If
we can realize a situation in which only two gravitons among the modes
in the Kaluza-Klein tower survive at low energies, bi-gravity
models may arise. In the braneworld setup, it would be easy to add
a bulk scalar field to the models in such a way that only a single
scalar mode effectively remains at low energies. However, in this
case, the kinetic term of this low-energy scalar field will necessarily
contain both metrics corresponding to the two low-energy gravitons.
By construction, the braneworld models seem to avoid the appearance
of ghost. However, the result in the present paper tells that the
generic form of the kinetic term of a scalar field leads to the appearance
of the BD ghost. This may suggest that there is a crucial difficulty in
realizing the idea that a completion of ghost-free bigravity models
might be obtained by extending it to the braneworld setup. However,
the actual coupling between the low-energy scalar and gravitons in
the model derived from the braneworld setup can be more complicated
than the one we have 
discussed in this paper. Hence, we may find an alternative
way of coupling between matter and two metrics that avoids the appearance
of the BD ghost.

\acknowledgments
This work was supported in part by the Grant-in-Aid for Scientific
Research (Nos. 24103006, 24103001 and 26287044). 
We also would like to mention that the discussion during 
the long-term YITP workshop: YITP-T-14-1 was useful to complete this work.

\end{document}